\documentclass[aps,twocolumn,showpacs,preprintnumbers,amsmath]{revtex4}
\usepackage{dcolumn}
\usepackage{bm}
\usepackage{graphicx}
\usepackage{bm}
\begin{document}
\title{Low temperature study of field induced antiferro-ferromagnetic transition in Pd doped FeRh}
\author{Pallavi Kushwaha, Archana Lakhani, R Rawat and P Chaddah}
\affiliation{UGC-DAE Consortium for Scientific Research\\University Campus, Khandwa Road\\
Indore-452001, India.}
\date{\today}

\begin{abstract}
The first order antiferromagnetic (AFM) to ferromagnetic (FM) transition in the functional material  $Fe_{49}(Rh_{0.93}Pd_{0.07})_{51}$ has been studied at low temperatures and high magnetic fields. We have addressed the non-monotonic variation of lower critical field required for FM to AFM transition. It is shown that critically slow dynamics of the transition dominates below 50 K. At low temperature and high magnetic field, state of the system depends on the measurement history resulting in tunable coexistence of AFM and FM phases. By following cooling and heating in unequal magnetic field (CHUF) protocol it is shown that equilibrium state at 6 Tesla magnetic field is AFM state. Glass like FM state at 6 T (obtained after cooling in 8 T) shows reentrant transition with increasing temperature; viz. devitrification to AFM state followed by melting to FM state.

\end{abstract}


\maketitle\section {Introduction}

FeRh and its nearby compositions have been subject of extensive theoretical and experimental studies due to their various interesting magnetic properties \cite{Bara, Gu, Thie, Ju, Kouv, Kouv1, Shir, Nava, Lomm, Alga, Moru}. As prepared FeRh order in fcc lattice, where Fe and Rh atoms are randomly distributed \cite{Chao}. With annealing, it order in CsCl type bcc structure, where Fe and Rh atoms occupy the corner and center positions of the cube, respectively. Magnetically, in the CsCl type bcc structure, FeRh shows a para to ferromagnetic (FM) transition around $\approx650-670$ K ($T_{C}$) \cite{Kouv, Kouv1}. Magnetic moment of Fe and Rh atom in the FM state are reported to be $\approx3.2\mu_{B}$ and $\approx0.9\mu_{B}$ respectively \cite{Shir}. With decreasing temperature, it shows a first order ferro (FM) to antiferromagnetic (AFM) transition. This transition is sensitive to FeRh composition and preparation condition, therefore there is a large variation in FM to AFM transition temperature ($T_N\approx 320-370$ K) reported by various groups \cite {Bara, Nava, Lomm, Alga}. In the AFM state, there is no magnetic moment on Rh and the magnetic structure is type II AFM, where ferromagnetic Fe layers (111) are coupled antiferromagnetically to each other \cite{Shir, Moru}. This FM to AFM transition can also be influenced by the substitution of transition metal at Fe as well as Rh site \cite{Bara, Kouv1, Bara1}. Depending upon doping element ( e.g. Ir, Pt, Pd, Ni etc.) and concentration, $T_N$ can be shifted (upward/downward) over a wide temperature range \cite{Kouv1}. The origin of FM to AFM transition in this system is still debated. Since this transition is accompanied with an abrupt change in lattice parameter and unit cell volume, Kittel exchange inversion model has been used to explain the transition \cite{Kitt}. However, this model fails to describe various features associated with this transition like non-monotonic variation of $T_N$ with x in case of $Fe_{49}(Rh_{1-x}Pd_x)_{51}$ \cite{Bara}, anomalous entropy change, vanishing of Rh moment etc \cite{Bara, Anna}. Heat capacity measurements show four times higher electronic contribution to heat capacity in FM state when compared to AFM state \cite{Bara, Tu, Ivar}. This led Annaorazov et. al.\cite{Anna} and Tu et. al \cite{Tu} to suggested that band structure modification as the origin of FM to AFM transition. On the other hand Chen et al. \cite{Chen} reported small difference in the optical conductivity of FM and AFM phases through their ellipsometric studies. According to them, the low temperature difference in heat capacity of AFM and FM state has magnetic origin rather than electronic origin. Density functional calculations of Gu et. al. \cite{Gu} attributed the AFM-FM transition to the magnon (spin-wave) excitations. Besides the origin of the transition, interest in this system also arises due to their potential for technological applications. This is because FM to AFM transition is accompanied with large change in magnetization, resistivity, volume etc. and transition can be influenced by magnetic field as well as pressure \cite{Bara, Anna, Alga, Raja, Wayn}. As a consequence giant magnetocaloric effect \cite{Anna, Anna1}, elastocaloric effect \cite{Anna}, giant magnetoresistance \cite{Alga, Bara1}, magnetostriction \cite{Rico, Ibar, Marq} etc. have been observed in this system. Multilayers of FeRh/FePt films have been shown to form exchange spring system, which opens up the possibility for thermally assisted magnetic recording media \cite{Thie}.  Recently observation of laser induced ultrafast switching between AFM and FM state on sub picoseconds timescale in FeRh has opened another area of investigation \cite{Ju,Thie1}.

In spite of extensive investigations in this system, there are limited studies on the AFM-FM phase coexistence. Most of these studies are focused around room temperature or above, which is closed to $T_N$ in the studied system. A detail magnetization (M) investigation of $Fe_{49}Rh_{51}$ thin films by Maat et. al. \cite{Maat} showed heterogeneous AFM to FM transition during warming irrespective of substrate. However FM to AFM transition upon cooling on c-axis sapphire substrate film suggested homogeneous nucleation and growth of AFM domain. This study also showed thermomagnetic irreversibility in M-H and M-T measurement arising due to supercooling and superheating associated with first order transition. MFM study of Yokoyama et. al. \cite{Yoko} showed inhomogeneous nucleation of FM domains at a micrometer length scale in $FeRh_{0.24}Pd_{0.76}$ and attributed to the internal stress caused by anisotropic lattice expansion at the transition. Manekar et al. \cite{Mane} have studied the evolution of FM state in superheated AFM state by MFM in $Fe_{52}Rh_{48}$. They showed the coexisting AFM and FM phase in the sub-micrometer length scale and nucleation and growth of FM phase coupled with topography on a time scale of $10^4$ sec. On the other hand studies at low temperatures are rare in this system. Studies on fcc structured nanoparticles of FeRh showed spin glass behavior in magnetization measurement \cite{Nava, Hern}. Baranov et al. \cite{Bara} have carried out detailed studies of AFM-FM transition with various transition metal doping down to $2$ K. In $(Fe_{0.965}Ni_{0.035})_{49} Rh_{51}$ and $Fe_{49}(Rh_{0.92}Pd_{0.08})_{51}$, where the $T_N$ is shifted to low temperature, they showed that critical field and hysteresis width follow $T^2$ and $T^{1/2}$ dependence, respectively.  Below $5$ K (for Ni) and $3.5$ K (for Pd), they have noticed  scattered but substantially lower hysteresis width ($\Delta H_c$) and upper critical field (field required for AFM to FM transition) than the extrapolated curve obtained from high temperature data. These features have been attributed to macroscopic quantum tunneling of magnetization through the energy barrier. Besides this, figure 2, 5 and 8 of Baranov et. al. \cite{Bara} also reveal non-monotonic variation of lower critical field showing a maxima at much higher temperature ($\approx 60$ K). Similar kind of non-monotonic variation of lower critical field has been shown and addressed in $Nd_{0.5}Sr_{0.5}MnO_3$ \cite{Rawa}, $Mn_{1.85}Co_{0.15}Sb$ \cite{Pall}. There, such anomalous behavior has been explained in terms of critically slow dynamics of the transition on measurement time scale, which gives rise to glass like arrested state (GLAS) at low temperature. In case of $La_{5/8-0.4}Pr_{0.4}Ca_{3/8}MnO_3$ (LPCMO) it has been shown that combination of glass transition temperature $T_G(H)$ and supercooling  line $T^*(H)$ gives rise to non-monotonic hysteretic boundary \cite{Wu}. GLAS arises out of kinetic arrest of a first order magnetic transition and is different from spin or cluster glass \cite{Roy, Roy1}. In analogy to structural glasses, GLAS shows devitrification on warming giving rise to large change in volume fraction of magnetic states \cite{Roy}. GLAS and phase coexistence are of current interest particularly in manganites \cite{Roy}. Therefore the study of first order magnetic transition in FeRh system provides an opportunity to check the universal feature of GLAS beyond manganites.
 
We have chosen $Fe_{49}(Rh_{0.93}Pd_{0.07})_{51}$ for the present study. This is because in $Fe_{49}(Rh_{1-x}Pd_{x})_{51}$ system, $T_N$ is reported to decrease from $\approx 350$ K ( x= 0) to $ \approx180$ K (x=0.08) and then increase again with increasing x before disappearing for $x\ge 0.13$ \cite{Bara}. Therefore $T_N$ is minimum around $x \approx 0.08$. We carried out detailed magnetotransport studies on $Fe_{49}(Rh_{0.93}Pd_{0.07})_{51}$ which shows that dynamics of the transition becomes critically slow at low temperature and high magnetic field resulting in a non-monotonic variation of lower critical field. It results in coexistence of glass like arrested FM state and AFM state at low temperature.  A CHUF (cooling and heating in unequal field) protocol \cite{Bane} has been used to find out the equilibrium state of the system. This is the first study where critically slow dynamics of the transition is observed for AFM-FM transition accompanied with creation or vanishing of magnetic moment (on Rh).

\maketitle\section{Experimental Details}

The compound $Fe_{49}(Rh_{1-x}Pd_{x})_{51}$ with $x=0.07$ was prepared by arc melting the constituent elements of purity better than $99.9\%$ under high purity argon atmosphere. Small pieces, cut from the same ingot, were wrapped in Tantulam foil and sealed in a quartz tube in $\approx10^{-6}$torr of vacuum and annealed at $900^0C$ for 20 hours. For crystal structure analysis and phase detection, X-ray diffraction has been performed on polished surface of the ingot. The resistivity measurements were carried out by standard four-probe technique using a home made resistivity setup with 8 Tesla superconducting magnet system from Oxford Instruments Inc., UK. All the in-field measurements were performed in longitudinal geometry except isothermal magnetoresistance (MR) measurements, which were performed  in transverse geometry up to 14 Tesla using Physical Property Measurement System (PPMS) from M/s. Quantum Design, USA. The magnetoresistance (MR) is defined as MR $= \left\{\rho (H)- \rho(0)\right\}/\rho(0)$, where  $\rho(0)$ is the resistivity in zero field and $\rho (H)$ is the resistivity in the presence of magnetic field \textsl{H}. Magnetization measurements were performed using Vibrating Sample Magnetometer (VSM) option of PPMS.

\maketitle\section{Results and discussion}

Figure 1 shows the X-ray diffraction pattern of as cast and annealed $Fe_{49}(Rh_{0.93}Pd_{0.07})_{51}$. As cast sample is indexed by considering the presence of both fcc (it is fct with lattice parameter a and c almost equal)  and bcc structure and  corresponding peaks in the figure are marked by stars (*) and vertical lines (\texttt{|}) respectively. After annealing at $900^{0}C$ for 20 hours, almost all of the fcc phase is converted into an ordered bcc structure. This observation is consistent with the earlier studies \cite{Kouv1, Ibar}.

\begin{figure}[t]
	\begin{center}
	\includegraphics[width=8 cm]{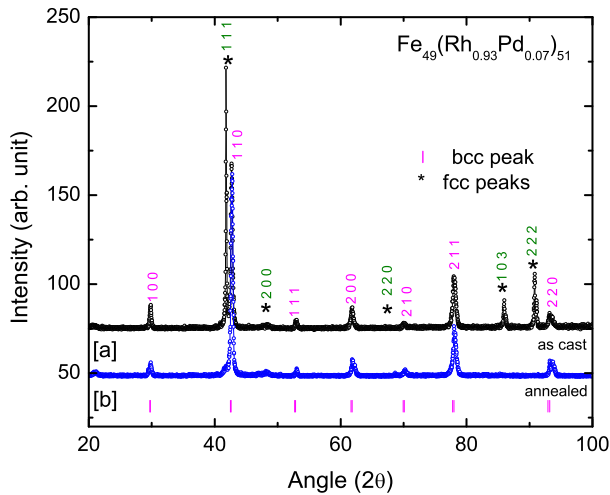}
	\end{center}
	\caption{ X-ray diffraction pattern for [a] as cast and [b] annealed sample. Peaks marked by stars (*) and vertical lines (\texttt{|}) correspond to fct and CsCl type bcc structure respectively.}
	\label{Figure1}
\end{figure}
 	
  \begin{figure}[t]
  \begin{center}
	\includegraphics[width=8 cm]{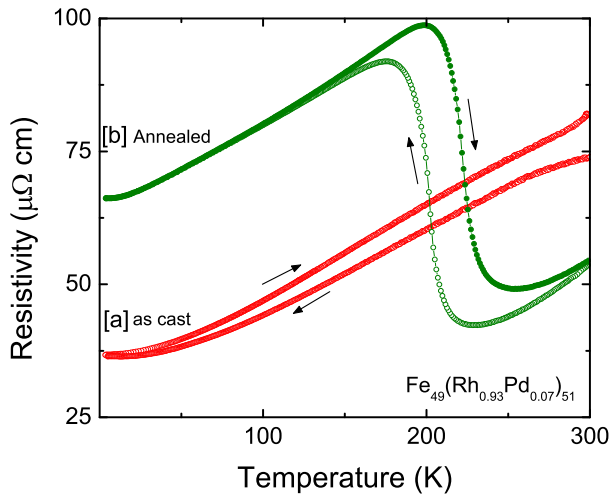}
	\end{center}
	\caption{Resistivity behavior as a function of temperature for [a] as cast and [b] annealed sample. A first order transition from FM (low resistivity) to AFM (high resistivity) state can be clearly seen in annealed sample.}
	\label{Figure2}
	\end{figure}	
	
 	Figure 2 shows the temperature dependence of resistivity for both the samples in the absence of magnetic field. As cast sample does not show any transition below room temperature but a difference between resistivity value during cooling and heating cycle is noticeable over entire temperature range. The origin of this hysteresis is not clear to us. This is similar to \textsl{M-T} measurement on FeRh filings by Lommel et al. \cite{Lomm}, where magnetization at 300 K is smaller after cooling the sample to 78 K and warming back to 300 K. Annealed sample shows a sharp rise in resistivity with decreasing temperature which indicates transition from low resistive FM phase to high resistive AFM phase. During heating, a reverse transformation from AFM to FM state occurs at higher temperature resulting in a hysteresis, which is a signature of first order nature of the transition. The transition temperature, taken as the inflection point of the resistivity curve, is found to be $\approx 201$ K during cooling and $\approx 222$ K during heating. Besides this, FM to AFM transition is quite broad as transition width during cooling as well as heating is found to be around 55 K. The studies around $T_N$ in analogous system has shown the presence of coexisting AFM and FM phase and broadening is attributed to residual lattice imperfection, chemical disorder along with the internal stress caused from anisotropic lattice expansion \cite{Maat, Mane, Yoko}. Due to disorder, different regions having length scale of the order of the correlation length can have different transition temperatures $T_N$, and this results in broadening of transition line as well as supercooling $(T^*,H^*)$/superheating $(T^{**},H^{**})$ spinodals into a band for a macroscopic sample \cite{Imry}. Since we are interested in the study of first order AFM-FM transition at low temperature, further studies are performed on annealed sample only, which will be discussed in the following sections.
 	 	
 	\begin{figure}[b]
	\begin{center}
	\includegraphics[width=8 cm]{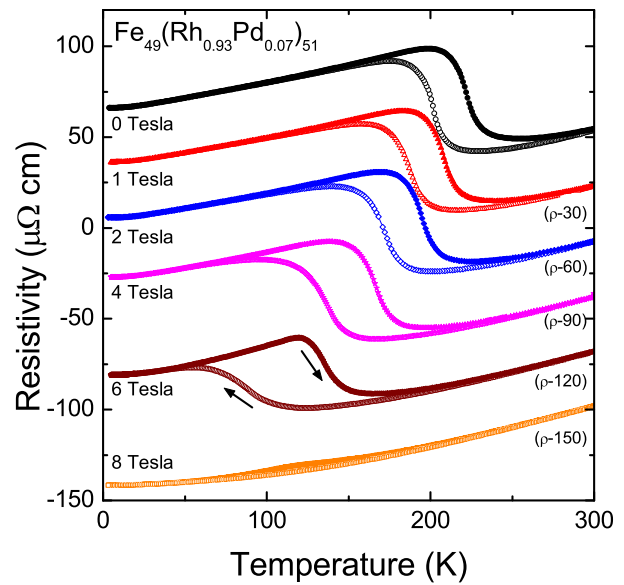}
	\end{center}
	\caption{Temperature dependent resistivity in different magnetic field conditions. Measurement has been performed during cooling (open symbol) and subsequent warming (solid symbol) in the presence of labeled magnetic field. Y-axis is for 0 Tesla curve. For the sake of simplicity, other curves are shifted downwards by labeled value e.g. label ($\rho -30$) indicates that resistivity curve is shifted by 30 $\mu\Omega$-cm.}
	\label{Figure03}
	\end{figure}	

Figure 3 shows the temperature dependence of the resistivity in the presence of various constant magnetic fields. Labeled magnetic fields are applied isothermally at 300K and data is recorded during cooling (FCC) and subsequent warming (FCW). Up to 4 T magnetic field, $T_N$ decreases linearly ($\approx 16$ K/T) and both transition width ($\approx 55$ K) as well as hysteresis width ($\approx 21$ K) remains almost constant. This observation of constant hysteresis width with varying magnetic field is in agreement with the studies on analogous systems having $T_N$ close to or above room temperature \cite{Maat}. For 6 T magnetic field, transition width during cooling increases significantly in spite of decrease in resistivity jump across the transition. Since $\rho$ value does not reach the zero field resistivity  value even after completion of hysteretic region, it suggest the presence of FM phase coexisting with AFM phase at low temperature. In the presence of 8 T magnetic field, there is no clear indication of FM-AFM transition, however, presence of small hysteresis indicates partial transformation of FM phase into AFM phase during cooling. These results show that with the application of magnetic field FM-AFM transition is suppressed only up to $\approx 50$ K (instead of reducing toward 0 K) and no FM to AFM transformation takes place below this temperature. It is worth noting here that even transition metal substitution studies show abrupt vanishing of AFM-FM transition temperature with substitution (or first order transition with composition) and $T_N$ is always found to be either higher than 100 K or absent \cite{Bara}.

	\begin{figure}[b]
	\begin{center}
	\includegraphics[width=8 cm]{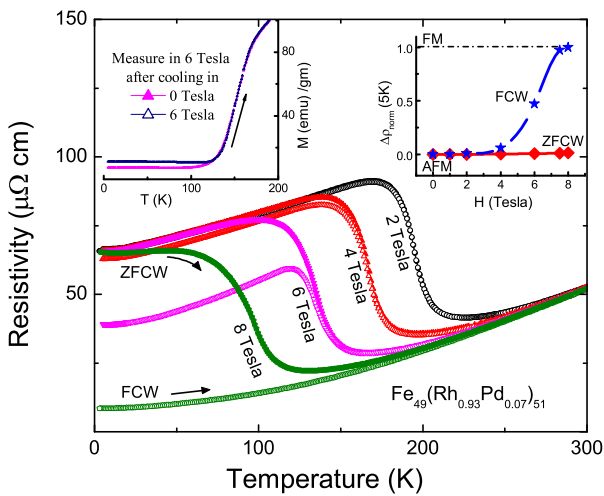}
	\end{center}
	\caption{Resistivity vs. temperature in the presence of labeled magnetic field measured during warming. ZFCW curves were measured after cooling in zero field and FCW curves were measured after cooling in same field value. The left  inset shows the temperature dependence of magnetization in the presence of 6 T magnetic field. The right inset shows $\Delta\rho_{norm}$ at 5 K for FCW (blue star) and ZFCW (red diamond) curve, highlighting the path dependence of FM and AFM phase fraction (see text for details).}
	\label{Figure04}
	\end{figure}

To verify if these co-existing states in 6 T and FM states in 8 T are ergodic state we followed different path for resistivity measurement under same temperature and magnetic field values as used in figure 3. Now, system is cooled to 5 K under zero field and labeled magnetic field applied isothermally at 5 K and resistivity is measured during warming (ZFCW). Figure 4 shows ZFCW curves along with corresponding FCW curves taken from figure 3. For 2 T and 4 T both the curves are almost identical. However for 6 T and 8 T, ZFCW curves have higher resistivity compared to FCW curve below $T_N$. Left inset of figure 4 shows the ZFCW and FCW magnetization curves in the presence of 6 T magnetic field. It shows higher magnetization for FCW compared to ZFCW. A difference between ZFCW and FCW in magnetization measurement has been observed by Navarro et. al. \cite{Nava} and Hernando et. al. \cite{Hern} on ball-milled fcc FeRh. There the difference between ZFCW and FCW decreases with increasing magnetic field and is attributed to spin glass behavior. However in our case, the difference between ZFCW and FCW increases with increasing magnetic field (but less than critical field required for AFM-FM transition at 5 K) which rules out the presence of spin glass. The difference between ZFCW and FCW curves reduces as the applied field becomes higher than 9 Tesla because the isothermal AFM to FM transition starts at 5 K, as shown in the R vs H curve in figure 5(a). Consequently, at 14 Tesla, system will be in FM state along ZFCW curve also. These thermo-magnetic irreversibility (TMI) for resistivity in ZFCW and FCW curves are similar to that observed across disorder broadened first-order magnetic transition in many other systems like doped $CeFe_2$ \cite{Roy}, $Nd_{0.5}Sr_{0.5}MnO_3$ \cite{Rawa}, $Mn_{1.85}Co_{0.15}Sb$ etc. \cite {Pall}. There it has been attributed to critically slow dynamics of the transition due to which high temperature phase remains arrested down to the lowest temperature. To have some estimate of the FM phase fraction at 5 K we assumed that resistivity decreases linearly with increasing FM phase fraction and the resistivity values corresponding to FM state ($\rho_{FM}$) and AFM state ($\rho_{AFM}$) are taken as the 8T FCW curve and  0T (zero field cooled) curve resistivities at 5 K, respectively. With this assumption the quantity $\Delta\rho_{norm}$ [$= \left\{\rho_{AFM}- \rho(H, T)\right\}$/$\left\{\rho_{AFM}-\rho_{FM}\right\}$] gives FM phase fraction at temperature $T$ and magnetic field $H$. Obtained $\Delta\rho_{norm}$ at 5 K is shown as inset in figure 4 for both ZFCW and FCW curve. For ZFCW curve $\Delta\rho_{norm}$ remains zero for all the field values which indicates that the system is in homogeneous AFM state at 5 K and applied magnetic field is not sufficient enough to induce AFM to FM transition. Whereas, in case of FCW curve it deviates from ZFCW curve for $H\geq 6$ T that shows increased FM phase fraction with increasing magnetic field. In case of 6 T, it indicates coexisting AFM and FM phase and the state of the system depends on the path followed in $H-T$ space.  On the other hand we could obtain almost homogeneous AFM or FM state in presence of 8 T magnetic field depending on the cooling history.

  \begin{figure}[t]
	\begin{center}
	\includegraphics[width=8.5 cm]{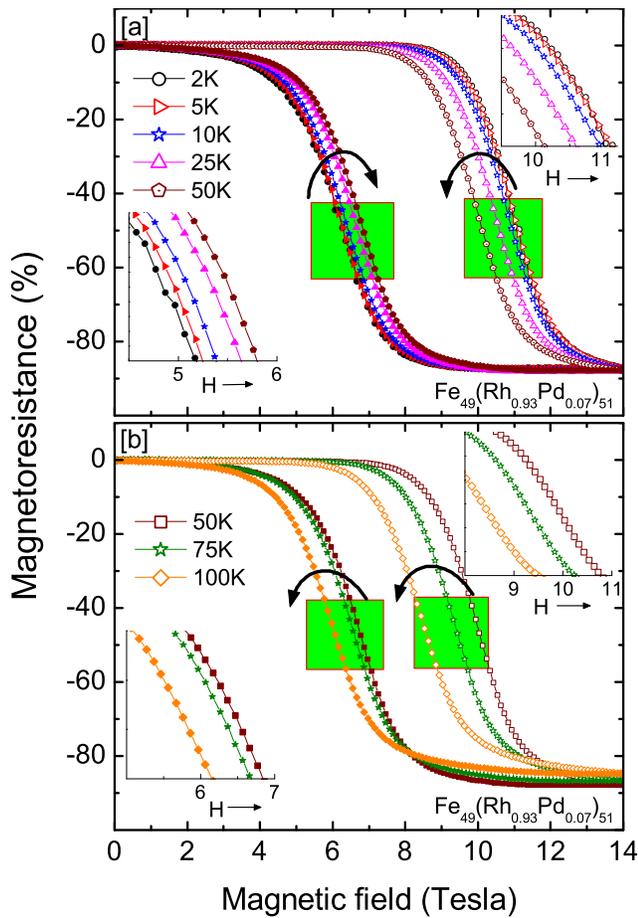}
	\end{center}
	\caption{The magnetic field dependence of resistivity at various constant temperatures are shown in [a] below 50 K and [b] above 50 K. Increasing and decreasing magnetic field cycles are denoted by open and solid symbols respectively. Upper right insets of both the graphs show the enlarged section of increasing field cycle whereas lower left inset correspond to decreasing field cycle. The forward curves ($0 \rightarrow 14$ T) move to lower field values monotonically with increasing temperature ($5 \rightarrow 100$ K) as highlighted by the curved arrows. Whereas, the reverse curves (14 T $\rightarrow 0$) move non-monotonically with temperature. Below/above 50 K, it shifts to higher/lower field value with increasing temperature as highlighted by oppositely directed curved arrow in figure [a] and [b] respectively.}
	\label{Figure05}
	\end{figure}
	
Since the isothermal application of 8 T magnetic field at 5 K does not show any field induced AFM-FM transition for a zero field cooled sample, we used PPMS for measurement up to 14 T to study this transition isothermally below $T_N$. These measurements were carried out in transverse geometry. We cross checked results of these measurements with the measurement of longitudinal geometry using 8 T magnet system. We found similar behavior in both the geometries except slightly higher MR magnitude ($\approx4\%$) and higher critical field (difference $\approx0.5$ T) in transverse geometry though the trend remains same. For these isothermal MR measurement, sample was cooled under zero field from well above $T_N$ to the lowest temperature (2 K) of the measurement. All the measurements have been done during subsequent warming in such a way that the every next higher temperature measurement was performed after completion of previous low temperature measurement. The results of these measurements are shown in figure 5. With the application of magnetic field, a field induced transition from AFM to FM phase is observed as a sharp change in MR. For all the temperatures, giant MR $\approx 90\%$ is observed across AFM to FM transition. The reverse transformation occurs at lower field value resulting in a hysteretic field dependence of MR. During decreasing magnetic field, system recovers its initial resistivity value at 0 T, which indicates complete reverse transformation from FM to AFM phase. This high MR is observed when system is prepared under zero field condition and magnetic field is cycled back to zero. Whereas in figure 4 it has been shown that if the sample is prepared in FC state same large MR would not be observed. The right insets of fig 5 (a) and (b) shows that with increasing temperature, forward ($0 \rightarrow 14$ T) curves shift monotonically to lower field side. However the return curves (14 T $\rightarrow 0$) shift to higher field side up to 50 K (left corner inset of fig.5(a)) and then to lower field side (left corner inset of fig.5(b)) with increasing temperature. This behavior is similar to that observed in $Nd_{0.5}Sr_{0.5}MnO_3$, where it is attributed to dominance of critically slow dynamics of the transition below 60 K \cite{Rawa}. 
    
 \begin{figure}[b]
	\begin{center}
	\includegraphics[width=8 cm]{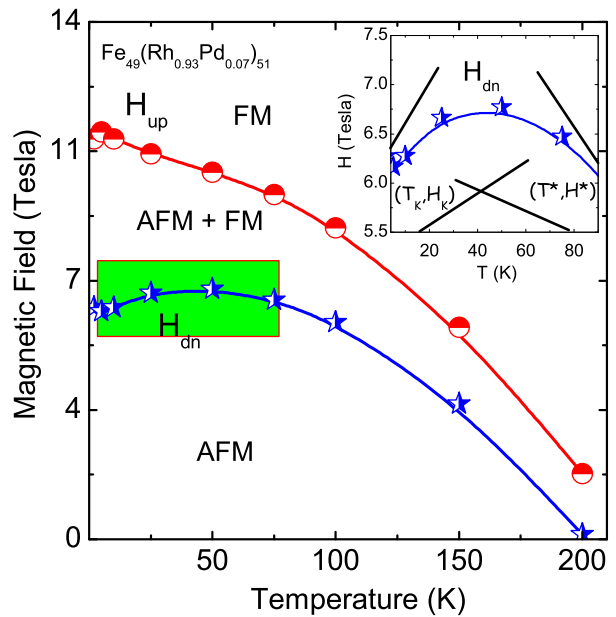}
	\end{center}
	\caption{$H-T$ phase diagram obtained from isothermal MR measurements shown in figure 5. Inset highlights the non-monotonic variation of lower critical field and straight lines represent $(H_K, T_K)$ and $(H^*, T^*)$ band schematically.}
	\label{Figure06}
	\end{figure}   
	
  The variation of critical fields is more explicitly demonstrated in $H-T$ phase diagram (figure 6) which is obtained from isothermal magnetoresistance shown in figure 5. $H_{up}$ (critical field required for AFM to FM transition) and $H_{dn}$ (critical field required for FM to AFM transition) are taken as inflection point of MR vs. $H$ curve during increasing and decreasing magnetic field respectively. This phase diagram shows that curve corresponding to $H_{up}$ varies monotonically with temperature, whereas $H_{dn}$ varies non-monotonically with a shallow maxima around 50 K (see inset of figure 6). Non-monotonic variation of lower critical field is addressed in $Nd_{0.5}Sr_{0.5}MnO_3$ \cite{Rawa} and $Mn_{1.85}Co_{0.15}Sb$ \cite{Pall} in terms of interplay between transformation kinetics and supercooling. This interplay is highlighted schematically in the inset of figure 6 for the present system. At low temperature (below $(H_K,T_K)$) dynamics of the transition becomes critically slow and wins over thermodynamic transition. Since dynamics of the transition dominates at low temperature, $H_{dn}$ is determined by $(H_K, T_K)$ in this temperature region, in contrast to high temperature where it is determined by $(H^*, T^*)$. Therefore the positive slope region is representative of $(H_K, T_K)$ and the negative slope region is representative of $(H^*,T^*)$. With this understanding of phase diagram it is obvious that transition field at which the free energy of both the FM and AFM state becomes equal will be higher than the middle point of $H_{up}$ and $H_{dn}$ at low temperatures. Nonmonotonic variation of critical field is common in manganites where such glassy magnetic states are being extensively studied. In case of LPCMO where high temperature state is AFM-charge order, Wu et al. \cite{Wu} have identified this kinetic arrest line ($H_K, T_K$) as $T_G$ (glass transition) line.

  \begin{figure}[t]
	\begin{center}
	\includegraphics[width=8 cm]{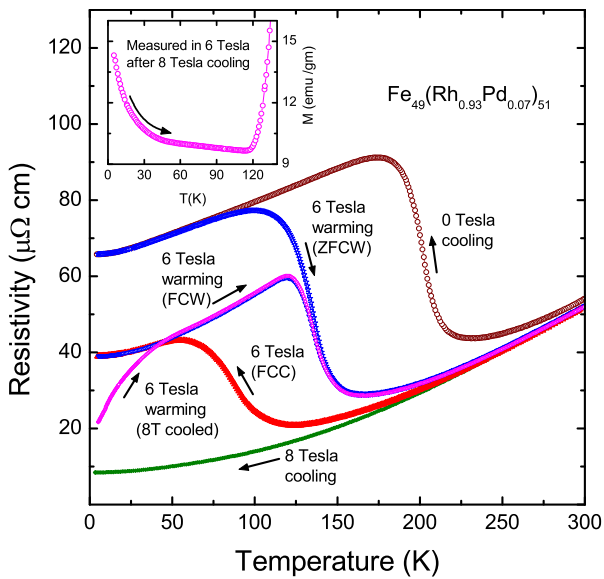}
	\end{center}
	\caption{Resistivity as a function of temperature using CHUF (cooling and heating in unequal field) protocol. Sample is cooled under different magnetic field of 0, 6 and 8 T respectively, whereas measurement during warming are carried out in 6 T only. A reverse transformation from arrested FM to AFM at low temperature is clearly seen during heating in 6 T after 8 T field cooling only. The usual first order AFM to FM transition is visible for all the warming curve for different field cooling. Inset shows the magnetization measured in the presence of 6 T magnetic field during warming after field cooling in the presence of 8 T magnetic field.}
	\label{Figure07}
	\end{figure}
	
So far we have shown that interplay of $(H_K, T_K)$ and $(H^*, T^*)$ give rise to coexisting AFM and FM phases at low temperature and the phase fraction depends on the path followed in $H-T$ phase space. Now question arises, which of these phases is the equilibrium phase. For this, CHUF (cooling and heating in unequal field) protocol \cite{Bane} is used, the results of which are shown in figure 7. Under this protocol, measurements during warming are carried out under a constant magnetic field after cooling the sample in the presence of different magnetic fields. If high field state happens to be non-equilibrium state then for cooling field higher than measuring field one observes a double (reentrant) transition and for smaller cooling field there will be only one transition. For the present system we cooled the system in the presence of 0, 6 and 8 T field to 5 K. At 5 K, magnetic field is changed isothermally to 6 T and then measurement is performed during warming in the presence of 6 T magnetic field. The resistivity value at 5 K and 6 T depends on the cooling field, which shows the tunability of AFM/FM phase fraction. Cooling in higher field results in higher FM phase fraction. When the cooling field (0 T) is lower then warming field (6 T), system shows only one transition i.e transformation from AFM to FM state around 125 K. Whereas two transitions appear when cooling field (8 T) is higher than measuring field (6 T). Corresponding magnetization curve (cooled in 8 T and measured in 6 T warming) in the inset of figure 7 also shows two transitions.  It shows that low field state (here AFM) is equilibrium phase and FM phase exist as glass like arrested phase. During warming in 6 T (after cooling in 8 T) an increase/decrease in the resistivity/magnetization at low temperature indicates devitrification of glass like FM phase into AFM phase. The re-entrant transition, corresponding to melting of AFM phase in to FM phase, is seen by a sharp fall/rise in resistivity/magnetization after crossing superheating band with further increase in the temperature.
  
These results show that cooling in high magnetic field results in glass like FM state as the FM state obtained in this way remains arrested until system crosses $(H_K, T_K)$ band on lowering the magnetic field isothermally (see figure 5). The origin of glass like arrested magnetic state is debatable even in manganites, where it has been subject matter of extensive investigation in recent years. Like manganites, there is a delicate balance between AF and FM interaction in FeRh system. This can be seen in figure 2, where we observe a sharp transition in annealed (chemically ordered) sample in contrast to broad features observed in the as cast sample. Not only this band structure in AFM and FM state are shown to be different. Therefore there are strong coupling between magnetic, electronic and lattice degrees of freedom and AFM-FM transition is sensitive to disorder, strain etc. In case of LPCMO Sharma et al. \cite{Roy} has linked the glass formation to freezing of structural degrees of freedom. According to them potential mechanism for glass formation lies in the first order structural phase transition. In the present system also AFM-FM transition is accompanied with large lattice volume change though the crystal structure symmetry remains same. The MFM study of Manekar et al. around room temperature in FeRh system has shown the growth of FM phase correlated with topology on a time scale of $10^4$ secs \cite{Mane}. We would like to caution here that the driving mechanism (lattice or magnetic) for first order transition in this system is yet to be identified. When the transition is shifted to lower temperature (in the present system achieved by Pd-doping and applying magnetic field) the atomic motion is hindered at low temperature due to lower thermal energy. The effect of Pd doping appears to be similar to metallic glass where melting temperature (but not glass transition temperature) is found to be strong function of composition \cite{Gree}. In fact even in the case of LPCMO thin film grown on different substrate, glass transition temperature appears to be less sensitive to substrate strain in comparison to AFM-FM transition temperature \cite{sathe}.

  \maketitle\section{Conclusions}
To conclude we have studied first order AFM to FM transition at low temperature using detailed magnetotransport studies on $Fe_{49}(Rh_{0.93}Pd_{0.07})_{51}$. Similar to abrupt vanishing of $T_N$ with transition metal doping \cite{Bara}, these studies show gradual decrease of transition temperature but only down to 50 K and absence of AFM-FM transition below this temperature. Non-monotonic variation of $H_{dn}$ observed in isothermal MR measurement has been explained by the interplay of $(H_K,T_K)$ and $(H^*, T^*)$ band. At low temperature and high magnetic field, state of the system depends on the measurement history and can give rise to coexisting AFM and FM state. Nature of coexisting AFM and FM phase has been studied by following novel paths in $H-T$ space which shows FM state as GLAS and its devitrification with increasing temperature. The observed glass like features in the studied system are similar to glassy magnetic states observed in manganites and related systems. The applicability of the concepts developed in these studies to the present system shows universality. Further microscopic studies will be helpful in understanding the glassy behavior in these systems.

\maketitle\section{Acknowledgments}
We thank N.P. Lalla, R.J. Choudhary and Suresh Bhardwaj for XRD measurements, V. Ganesan and Swati Pandya for isothermal magnetoresistance measurements, Alok Banerjee and Kranti Kumar for magnetization measurements. DST, government of India is acknowledged for funding the PPMS and VSM-PPMS used in the present study. Pallavi Kushwaha acknowledges CSIR, India for Senior Research Fellowship.

\end{document}